\documentclass{rpioaj}
\usepackage{cite}
\usepackage{amsmath,amssymb,amsfonts}
\usepackage{algorithmic}
\usepackage{graphicx}
\usepackage{textcomp}
\usepackage{setspace}

\usepackage{multirow}

\usepackage{booktabs}

\usepackage{pifont}

\graphicspath{ {figs/} }

\def\BibTeX{{\rm B\kern-.05em{\sc i\kern-.025em b}\kern-.08em
    T\kern-.1667em\lower.7ex\hbox{E}\kern-.125emX}}
\begin{document}
\history{\textbf{Received}: 1 April 2020; \textbf{Revised}: 7 Jun 2020; \textbf{Accepted}: 8 Jun 2020; \textbf{Published Online}: 10 June 2020\\ 
\textit{Researchpedia Journal of Computing, Volume 1, Issue 1, Article 6, Pages 50--65, Jun 2020}}
\doi{10.1109/RpJC.2020.DOI Number}

\title{Issues and challenges in Cloud Storage Architecture: A Survey}
\author{Anwar Ghani\authorrefmark{1}, 
Afzal Badshah\authorrefmark{1}, Saeed~Ullah~Jan\authorrefmark{2},
Abdulrahman A. Alshdadi\authorrefmark{3}
and~Ali~Daud\authorrefmark{3}
}
\address[1]{Department of Computer Science \& Software Engineering, International Islamic University Islamabad, 44000, Pakistan (e-mail: anwar.ghani@iiu.edu.pk, afzal.phdcs120@iiu.edu.pk)}

\address[2]{Department of Computer Science \& IT, University of Malakand, Chakdara, 18800, Pakistan (e-mail:saeedullah@uom.edu.pk)}

\address[3]{Department of Information Systems and Technology, College of Computer Science and Engineering, University of Jeddah, Saudi Arabia (e-mail: alshdadi@uj.edu.sa, ali\_msdb@hotmail.com)}


\markboth
{Ghani: \headeretal: Issues and challenges in Cloud Storage Architecture: A Survey}
{Ghani: \headeretal: Issues and challenges in Cloud Storage Architecture: A Survey}

\corresp{Corresponding author: Anwar Ghani(e-mail: anwar.ghani@iiu.edu.pk).}
\setcounter{page}{50}

\begin{abstract}

From home appliances to industrial enterprises, the \emph{Information and Communication Technology (ICT)} industry is revolutionizing the world. We are witnessing the emergence of new technologies (e.g, Cloud computing, Fog computing, Internet of Things (IoT), Artificial Intelligence (AI) and Block-chain) which proves the growing use of ICT (e,g. business, education, health and home appliances), resulting in massive data generation. It is expected that more than 175 ZB data will be processed annually by 75 billion devices by 2025.
The \emph{5G technology} (i.e. mobile communication technology) dramatically increases network speed, enabling users to upload ultra high definition videos in real-time, will generate a massive stream of big data. Furthermore, smart devices, having artificial intelligence, will act like a human being (e.g, a self-driving vehicle etc) on the network, will also generate big data. This sudden shift and massive data generation created serious challenges in storing and managing heterogeneous data at such a large scale. 
This article presents a state-of-the-art review of the issues and challenges involved in storing heterogeneous big data, their countermeasures (i.e, from security and management perspectives), and future opportunities of cloud storage. These challenges are reviewed in detail and new dynamics for researchers in the field of cloud storage are discovered.

\end{abstract}


\titlepgskip=-15pt

\maketitle

\section{INTRODUCTION}

The recent advances and development in smart technology is getting more  attention and attraction, resulting in a \emph{massive data} generation. The 75~billion devices forecast is a big number; even ten  times greater than the whole world population \cite{Prop149}. These devices  will increase the annual size of the global data-sphere up to 175~ZB \cite{Prop17,Prop154}. Another report states, as shown in Fig.~\ref{fig:forecost}, that more than 331 billion dollars will be invested in cloud up to 2023 \cite{Prop17}. This development not only requires special infrastructural improvements but also special techniques to process and store the incoming data \cite{Prop04}. Furthermore, integrating Artificial Intelligence (AI) in smart devices makes the data storage process more complicated. This big data expectation increases the need for cloud storage and related challenges to be explored.  
Fig.~\ref{fig:forecost} shows the devices and revenue forecast of cloud computing \cite{Prop17}. 

\Figure[!t](topskip=0pt, botskip=0pt, midskip=0pt)[width=0.45\textwidth]{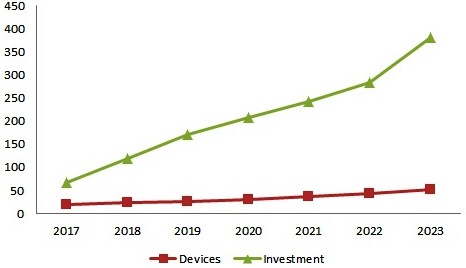}
{The cloud devices and revenue forecost.\label{fig:forecost}}
 

\Figure[!t](topskip=0pt, botskip=0pt, midskip=0pt)[width=\textwidth]{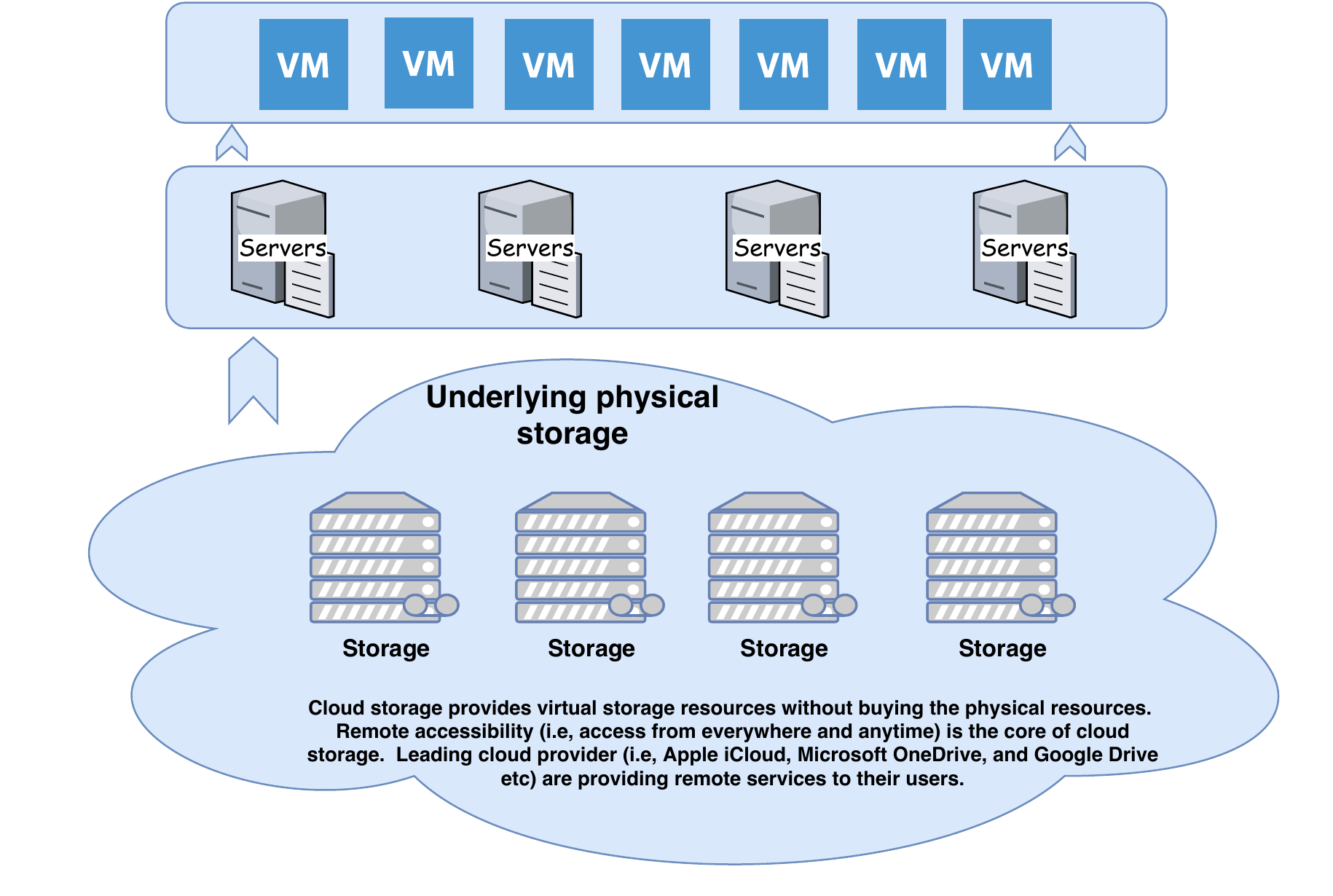}
{Structure of cloud storage.\label{fig:storage-architecture}}


Today more \emph{embedded devices} joined the Internet to monitor and connect everything (e.g, traffic facilities, buildings, environment, and lakes), enlarging the size of the data generation  \cite{Prop171, Prop172, Prop173, Prop174, Prop175}. As the data on the Internet is increasing day by day, therefore, analyzing and storing it through traditional data management method is a great challenge \cite{Prop176, Prop177} . However, researchers are struggling to design new kinds of databases based on NoSQL for handling unstructured data at such a large scale \cite{Prop178, Prop179, Prop180}. There are many proposals in the literature for a universal storage architecture which supports multiple data models at the same time and can store big heterogeneous data in the cloud environment  \cite{p03,p04,p05,p29}.
Fig. \ref{fig:storage-architecture} shows the structure of cloud storage.

With the advent of technology, computing requirements of organizations grew exponentially prompting the organization to incorporate more computing and storage resources \cite{p50}\cite{Prop03}. Setting up systems at such large scale require more efforts and heavy investments prompting the enterprise customers to outsource their computing and storage resources \cite{Prop181, Prop182, Prop183, Prop184}. The users have no full control over the computing resources available through cloud over the Internet \cite{p51,Prop186}. Storage in the cloud is becoming a hot research venue today because new applications are data intensive which doubles storage capacity requirement as well as data usage every year. It prompted some commercial organizations to work for another cloud service called as ``on demand storage''. Currently, the storage providers are fixated towards other aspects related to cloud storage like cost issues, performance issues and incorporating multiple storage \cite{p31}\cite{Prop185}\cite{Prop05}\cite{Prop187}.
Fig. \ref{fig:master-data-node} shows the structure of master and data node in cloud storage architecture. 

\Figure[!t](topskip=0pt, botskip=0pt, midskip=0pt)[width=0.8\textwidth]{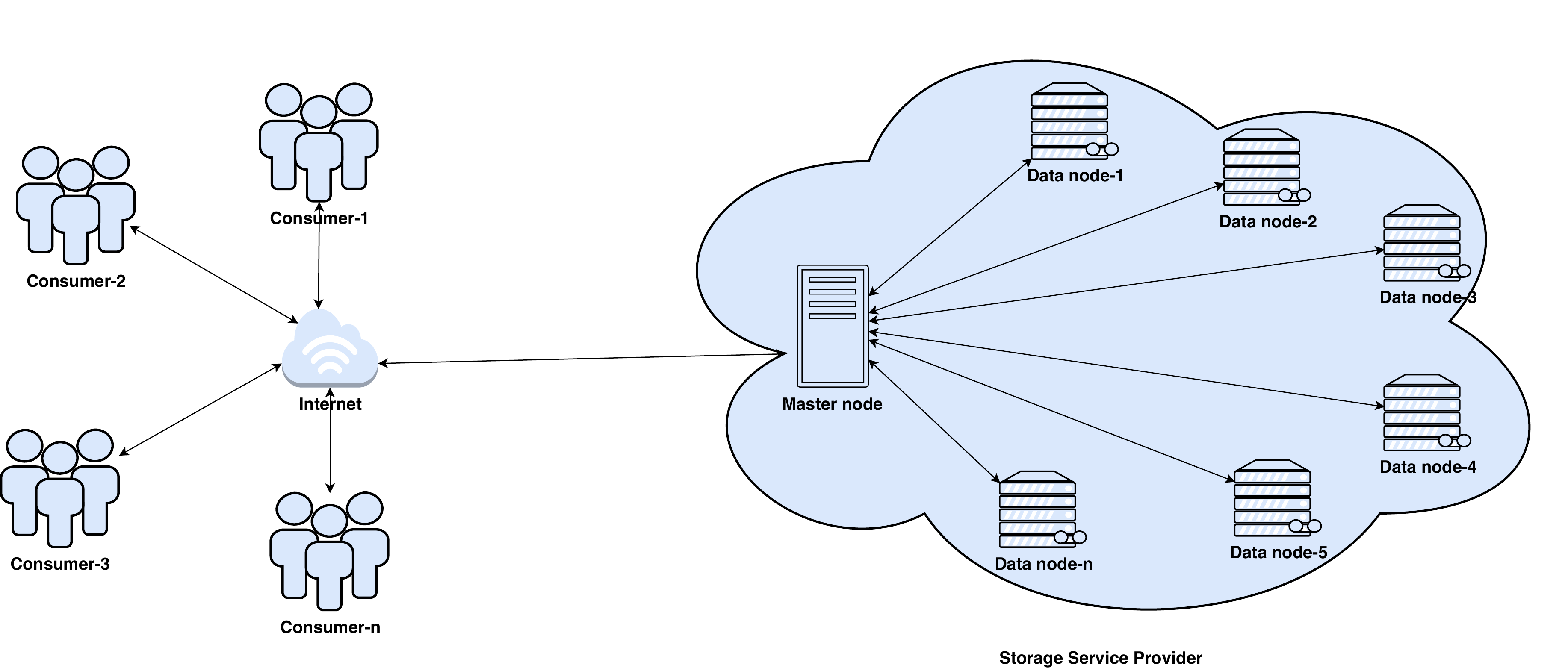}
{Master and data node in cloud storage architecture.\label{fig:master-data-node}}


The models of data centers in cloud computing are based on ``design-for-failure'' principle. Provisioning of global storage services require, cloud storage must use scalable, cheaper and purposed built solutions. Such solutions may include different hardware like, servers, networking equipment, and storage systems. It should use standard delivery models on massive economies of scale. ``off-the-shelf'' products designed for the traditional IT market may not be suitable to use in cloud data centers since they are not only expensive but also they do not meet the specific requirements of cloud data center environment.

This study explores the \emph{cloud storage architecture} its \emph{challenges} and \emph{possible solutions}. Additionally the  \emph{cloud storage future} and \emph{opportunities}. 
Cloud storage issues include but not limited to Security, Confidentiality, Data Dynamics, Integrity, Data Access, Data Segregation, Authentication and Authorization, Data Breaches, Backup Problem and vulnerabilities in Virtualization.

Rest of this article is structured as follows:
Section~\ref{cldStrg} provides an insight into the issues related to cloud storage and their countermeasures. 
Section~\ref{cldStrg} discusses the future opportunities of cloud storage. 
Finally, section~\ref{conclu} concludes the article with the key findings and future directions.


\section{Cloud storage challenges and possible solutions}
\label{cldStrg}

Storage in a cloud is a crucial part of the Infrastructure as a Service (IaaS). The lack of proper storage management in cloud environment, may lead to severe consequences \cite{p72}. Cloud storage related issues have been categorized as \emph{data security}  and \emph{data management} issues \cite{p19,p23}. This paper focuses on issues related to these two categories and a review of possible solutions to such issues.  Some of the points may overlap both categories, however, this distinction may help in understanding the challenges faced by cloud storage providers and tenants.  Fig \ref{fig:challanges} shows the challenges in cloud storage architecture. 
The following subsections elaborate these issues and their counter measures.

\Figure[!t](topskip=0pt, botskip=0pt, midskip=0pt)[width=0.8\textwidth]{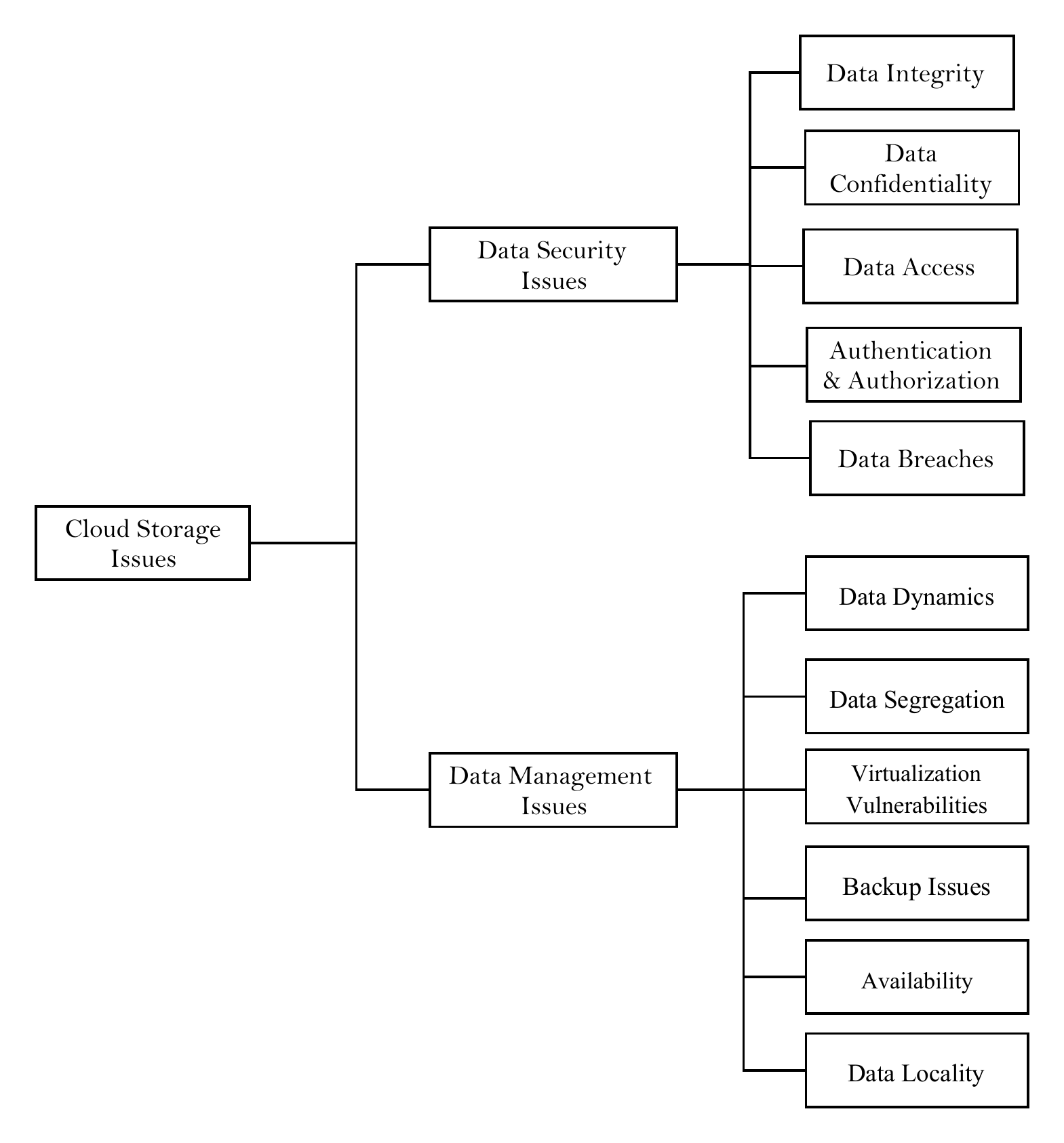}
{Cloud storage challenges.\label{fig:challanges}}


\subsection{Data Security Issues}

\emph{Data security} is an important requirement from tenant as a right.  Secure services attract users to store their data in a cloud. Companies providing the cloud storage services are searching for techniques that can control access to cloud data and improve security. With increase in size of the data, there is also an increase in data attacks and interceptions. The cloud computing provides storage services as a vitalized environment where a user has no control over the data \cite{p71}. In such situation, a user may ask questions like ``where exactly is my data located?'', ``what happen if I delete my data?'' and ``is the deleted data really deleted?''. 

\emph{Many solutions} to data security in cloud can be found in literature. Authors in \cite{p71} divided the security solutions into four layers (i.e. availability, authentication, confidentiality and integrity). They  argued that if confidentiality is achieved, it automatically ensures integrity. However, this sub section is dedicated to a more elaborate study of the issues related to data security.  
A recent study exploring data security and privacy in cloud storage \cite{p72} pointed out the three main reasons based on the features of cloud computing independent of the technology being used on the server. It includes outsourcing and  multitenancy.

A Time Stamp Authority (TSA) and Public Key Infrastructure (PKI) technologies are introduced into the cloud storage system for authentication and security with minimum cost and less system overhead. Trusted time stamp helps in audit and recording \cite{p45,p50,p52,p56}. The three points considered are User Identification, Time Stamping and User Verification through cloud storage system. The use of PKI improves security whereas authentication is done through directory services. The use of a time stamp provides security services like audit and evidences with a very minimum overhead. TSA also performs data management and optimization in cloud storage system. The workload is increased by TSA and client communication and the verification of users' operations. As during the communication process no certificate is used so extra overhead is not involved. The operation commands are converted into time stamp and sent to TSA server, which communicates with directory server and verify certificate. On validating the certificate, a time stamp is issued. The corresponding time stamp is then sent to the cloud and further operations may be performed. The cloud system stores the time stamps and operations record. The operations may be queries, downloads and uploads etc. The basic approaches used in designing data security techniques are shown in Table~\ref{secTec}. Furthermore, AI, 5G, IoT and block chain are improving the privacy and security \cite{Prop161}. 

\begin{table*}
	\centering
	\caption{Basic approaches used in designing data security techniques}
	\begin{tabular}{l|l l}
	\toprule
	\multirow{3}{*}{Data Security}   & Public Key Inscription & Low cost/system overhead\\
	\cmidrule(r){2-3}
	 & Trusted Timestamps	       & Auditing, recording, data management\\
	\cmidrule(r){2-3}			
	& Directory Services	       & Authentication, verification\\
	\bottomrule
	\end{tabular}
	\label{secTec}
\end{table*}

\subsubsection{Confidentiality Issues}

Cloud storage is a collection of storage servers on which multiple customers' data is stored, which makes privacy a major concern. The fundamental requirement for confidentiality of the information stored or processed in the cloud is the guaranteed protection of confidential or sensitive information. Based on the requirements of a specific scenario, this may relate to all or part of the externally stored data, the identity of the users who have access to the data or the actions that the users take on the data \ cite { t05}. Encryption techniques are used to achieve confidentiality in such systems. Cloud computing is a technology that uses the internet and servers to maintain and manage data and applications. Cloud computing has improved computing capabilities without large investments.

In the existing situation in order to avoid confidentiality issues, the system may want to implement encryption and decryption techniques \cite{p54} which lead to limited system operations and the user must know encryption decryption Keys. Some systems may implement both encryption and obfuscation depending on the type of data to be stored\cite{p71}.

A system based on proxy encryption, which supports various functions during the distributed storage system, is proposed in \ cite {p38, p55}, which consists of four stages: 1) system configuration, 2) data storage, 3) data transfer and 4) data recovery. An RSA-based algorithm is used to generate keys. The solution is when a sender `` A '' wants to send a message to recipient `` B '', `` A '' signs the message with his secret key and then encrypts it with the public key of `` B '' and downloads the encoded text. After retrieving the message, `` B '' decrypts it with its public key and then checks the public key sign `` A ''. The whole process involves two communication stages; a download from the sender `` A '' and download by the recipient `` B ''. This is why the proxy recording scheme is used to reduce the overhead of the data transfer function in the secure storage system. Here are some crucial points regarding data privacy in a cloud storage environment.

\begin{enumerate}
	\item In a cloud computing paradigm confidentiality of governmental and business information as well as privacy of personal information has the highest insinuations.
	
	\item The level of confidentiality and privacy of a user depends upon the privacy policies and terms of service provided by a cloud provider.
	
	\item Disclosure of information to a cloud provider by a user may change information of some specific types as well as certain user categories, rights and obligations of privacy and confidentiality.
	
	\item Personal and business information may be adversely affected in terms of legal status protection.
	
	\item Protecting confidentiality and privacy and the privacy rights of those processing and storing this information in a cloud environment may be highly affected by the location of information. 
	
	\item A cloud may store information at different venues with different legal implications leading to different legal consequences at the same time.
	
	\item Different laws against criminal activities and other matters can oblige/force a provider to disclose or examine user records for the sack of evidence.
	
	\item In addition to the legal protection for protecting a user's privacy and confidentiality, various legal qualms resist against gauging an information in a cloud for its status.
\end{enumerate}

\subsubsection{Integrity Issues}
\emph{Data integrity}  is one of the most crucial elements of any system. Integrity requires that the authenticity of the parties (i.e. users and vendors) communicating in the cloud guarantee the data stored with third-party vendors and the responses resulting from the calculation of requests \ cite {t05}. In a standalone system, data integrity may be achieved with a single database using constraints and transactions. To insure integrity of the data transactions must adhere the mostly used property in databases known as the ACID (atomicity, consistency, isolation and durability) property. But distributed systems are entirely different in complexity where multiple databases and multiple applications execution is a normal trait. In a distributed environment, data may be maintained at different sites. Therefore, any transaction involving data shared by multiple sites must be handled carefully in a way to avoid transaction failure and allow various distributed applications through a resource manager to be a part of the global transaction.

With the entrance to the world of Service Oriented Architecture (SOA) and Cloud computing, issues of data integrity grow exponentially because a mixture of local and  SaaS  (Software as a Service) applications are displayed as a service. SaaS model supports multi tenancy in applications which usually hosted by third party and their functionality is exposed through XML based APIs (Application Programming Interface). Similarly in other environments like SOA various applications uses web services for example SOAP and REST to expose their functionality. However, managing transactions using web services is a serious challenge. Since guaranteed delivery or transactions are not supported by HTTP  protocol level giving the only way out of implementing these SOA at the API level.

\subsubsection{Data Access Issues}
\emph{Issues in access} to data in a cloud storage are mostly due to security policies. For example, a small level business organization may use services of a cloud provider for executing its business processes \cite{p18,p36}. Such organizations allow their employees to access a specific organizational data according to its own organizational security policies. These policies may prevent some employees from accessing a specific set of data and allow them to access certain data. To stop intruders from gaining unauthorized access to cloud resources, a cloud must adhere these security policies \cite{p33}. The SaaS model must have the ability to allow organizations to integrate their security policies as well as keep organizational data within its boundary in case when multiple organizations use the same cloud environment. The requirement of availability is; there must be a mechanism for verification of Service Level Agreements (SLA) between a user and providers which verifies that the user's requirements are fulfilled \cite{t05}.   

Many counter measures proposed in the literature can be found to mitigate the problems related to data access in cloud storage. In literature three categories of secure access control can be found (i.e. Role Based Access Control (RBAC), User Based Access Control (UBAC) and Attribute Based Access Control (ABAC)) \cite{p37,p41}. Due to the attachment of access control list (ACL) to user data, UBAC is usually not considered as a suitable candidate for cloud storage. Additionally, the involvement of Big Data the computational and communication overhead required for handling ACL is high \cite{p40}. Then there is role based classification of users to control access to data. A user matching a specific role is granted access to data. Such approaches are considered suitable for business organizations at enterprise level for example hospitals \cite{p42}. The third and often used category in cloud storage is Attribute Based Access Control (ABAC) where a data owner assigns attributes and policies to users and data respectively \cite{p44}. In this case access to the data is granted to users having attributes that satisfy a specific access policy. For a confidential fine-grained access to data in cloud, this category is further divided into two approaches i.e. KP-ABE \cite{p43,p45} and CP-ABE \cite{p46}[13]. In case of KP-ABE the key of a user is linked with an access policy whereas attributes are linked with ciphertext. In contrast to KP-ABE, in CP-ABE the key of a user is linked with an attribute whereas the ciphertext is linked with an access policy.

However, the complexity of attribute based access control techniques grows linearly, as the number of attributes used in decryption raises, incorporating tremendous overhead in computation specially for devices with limited resources like mobile devices \cite{p36}.

\subsubsection{Authentication and Authorization Issues}

Authentication, in any system that needs a foolproof security, plays a crucial role like an entrance door that allows only trusted individuals, to the premises of a cloud. Access to important information depends on authentication, therefore, due to it's sensitive nature, authentication process must be robust to ensure availability to authentic users. In combination with cryptography, not only data confidentiality, but also its integrity  can be ensured by granting access only to authenticated individuals. Most of the security concerns can be mitigated through a sophisticated authentication mechanism \cite{p47,p48,p54}.

A Lightweight Directory Access Protocol (LDAP) server is used by various companies to store information about their employees \cite{p17}. Managing users in small and medium size businesses is mostly achieved  through Active Directory in the portion of business where the adoption of SaaS model is high (Microsoft White Paper, 2000). This model allows software to be hosted outside the organizational firewall. Many organizations separate user credential database from their IT infrastructure therefore, a customer must keep track of all the employees joining or leaving the organization and must enable or remove their accounts accordingly from the system. This may result in extra management overhead on the customer organization if it uses multiple SaaS products. In such cases different powers can be delegated to the customer by the provider, authentication for example enabling customer organizations internal LDAP/AD server to control their user management.

\subsubsection{Data Breaches}

A cloud environment is usually shared among many customers to store their data. Therefore, a compromise of the cloud environment means a potential threat to the data of all users making cloud an attractive target for attackers \cite{p18}. R. Cooper in his report \cite{m03} rated external criminals as the highest threat contributing 73\% but with least impact compromising 30,000 records producing 67,500 Pseudo Risk Score (PRS). Similarly, insider threats received the minimum rating of (18\%) but with greatest impact compromising 375,000 records with a PRS of 67,500. The middle rating has been received by partners with 73.39\% compromising 187,500 with a PRS of 73,125. The security provided by SaaS is argued to be better in comparison to conventional means, however insiders may not have direct database access but it still raises a risk with huge impact on data security. Employees of SaaS providers can cause exposure of customers private information since they have access to a lot of information. In order to avoid such complications, standards like PCI-DSS (Payment Card Industry-Data Security Standards) must be followed by SaaS providers.

\begin{table*}
	\centering
	\caption{Security solution for cloud storage architecture (Part-I)}
	\begin{tabular}{@{}p{3cm } |p{4cm} p{9cm}@{}}
	\toprule
	\textbf{Security Properties} & \textbf{Approaches} &  \textbf{Description} \\ 
	\midrule
	\multirow{2}{*}{Confidentiality} 
    & Cryptography  & Cryptography secure and protect data during communication. It is useful to block an unauthorized users from accessing private data. \\
	& Digital signatures	 &	Digital signature is a symbolic description that can verify the authenticity of messages or digital documents. A true digital signature provides access to the data.	\\
  & Proxy Re-encryption	 & Proxy encryption is commonly used when a party wants to reveal to a third party the content of messages sent to it that are encrypted with their public key. \\
    & Obfuscation        	 &  Obfuscation is the intentional creation of source code or machine code that is difficult for a human to understand. Like natural language eclipse, it can use unnecessarily redirected expressions to make statements. \\
    &  Blockchain        	 &  Blockchain is a smart design that offers digital information for sharing, but not for copying. Blockchain technology has created the backbone of a new type of internet. \\ 
    \midrule
	\multirow{4}{*}{Atomicity} 
    & MC                           &  Data MC is a world leader in the delivery of highly complex data migrations, specializing in end-to-end delivery of industry-specific, custom and enterprise transformation ERP, CRM projects. \\
	& Consistency                  &   The consistency of the database system refers to the fact that the database transaction can only be modified in an authorized manner. \\
	& Isolation                    & Isolation in database systems determines, how the integrity of activities are visible to other users and systems.\\
	& Durability                   &        In database systems, sustainability is the ACID property that ensures that closed transactions persist. \\
    \midrule
    \multirow{3}{*}{Data Access} 
    &   Role Based Access Control   &
RBAC is an entrance approach to regulate access to the system to authorized users. It is used by most companies with more than 500 employees and can implement mandatory access control (MAC) or discretionary access control (DAC). \\
	& User Based Access Control       & Role-based access (or role-based permissions) adds another categorization layer in addition to what is provided by user-based access. \\
	& Attribute Based Access Control  &  Attribute-based access control (ABAC) is  also called  as policy-based access control, defines an access control  that give access rights to users through the use of policies that combine attributes.        \\
	\midrule
    Data Breaches & Directory Services    & A Lightweight Directory Access Protocol (LDAP) server is used to provide authentication and authorization services \\
	\bottomrule
	\end{tabular}
	\label{secSol}
\end{table*}

\subsection{Data Management Issues}

The management issues related to data has been explored in this sub sections. The data management issues has been categorized and briefly explained as follows. 

\subsubsection{Data Dynamics Issues}

\emph{Data management}  in cloud is considered to be untrustworthy due to the fact that it shifts databases as well as application software to large centralized data centers. This new paradigm introduces various security issues yet to be understood. Data dynamics support through operations in cloud for example insertion, block modification, and deletion is a huge step in the direction of practicality as cloud services are not restricted only to backup and archiving. The following different methods are used for the assurance of data dynamics in cloud storage \cite{p57,p58}.

\begin{itemize}
	\item On a large scale the data centers are being transformed into computing pools by ``Software as a Service'' \emph{(SaaS)} computing architecture. In addition the fast growth in network resources like bandwidth and reliability enables customers to subscribe services with high quality from the remote data and software applications in data centers. 
	\item A cloud service provider for his own benefits may conceal errors in data or software used by the clients. For example a provider may deliberately delete data of an ordinary client which is accessed less often without the client's knowledge in order to increase his savings in money and storage space \cite{p57}. 
	\item For data dynamics various schemes have been designed with the efforts to combine efficiency, unlimited use of queries and information retrievablity  in these schemes. 
\end{itemize}

One possible solution in this case could be to motivate public auditing system of data storage security in Cloud computing \cite{p56}. In addition, fully dynamic protocols for data operations specially for block insertion, must be designed which is a lacking feature in most of the existing approaches. To support public auditing which is efficient and scalable, the existing schemes must be extended. Such extension should achieve batch auditing enabling a third party auditor (TPA) to perform auditing tasks delegated from multiple users simultaneously. 

\subsubsection{Data Segregation Issues}

Cloud computing architecture became popular because of it multi-tenancy nature \cite{p17,p59}. Multi-tenancy in cloud through SaaS applications allow storage of data from multiple users simultaneously. This may create an opportunity for a user's data to intrude into another user's data since data of different users reside at single location. This intrusion may exploit application's loopholes or by injecting SaaS system with malicious client code. If an application injected with a masked code executes it without verification shows that there are high possibilities of intrusion into others data. Therefore, a SaaS model must ensure that the data of each user is bounded both at physical and application levels. Data from different users must be ghettoise intelligently by the SaaS service \cite{p10}. 

Security checks may be bypassed using vulnerabilities in application by attackers through handcraft parameters. This may lead to the exposure of other tenants sensitive data. Therefore different assessments test must be performed to ensure that data from different users in multi-tenant environment is fully segregated from each other. These tests include; i) Data validation, ii) SQL injection flaws and iii) Storage insecurity. Any possible flaws detected by these tests could be used to illegally access sensitive data of the enterprise or other tenants.  

\subsubsection{Virtualization Issues and Vulnerability}

One of the major component of cloud environment which ensures that various instances running over a single machine be ghettoise from each other is known as virtualization. It is the source of major security challenges in a cloud environment which are not fully investigated today \cite{p15,p60}. Second issue is the administrative control of the operating systems, operating as guest and host systems and their imperfect provisioning of isolation \cite{p61} and scalability issues \cite{p63}. Many of the current Virtual Machine Monitors (VMM's) suffer from bugs allowing escape from VM therefore, ``root security'' is mandatory in such cases to prevent host operating system from being interfere with by any virtualized guest systems. Some virtualization software has been reported to have vulnerabilities which could allow a local user or an attacker to skip certain security checks and gain illegitimate access \cite{p61,p62}. One such example is that of Microsoft Virtual Server and Virtual PC vulnerability where a user of guest operating system could be allowed to execute code on other guest operating system or even the host operating system itself. This could allow a raise in privileges which can lead to unauthorized access of sensitive information. Similarly a validation error in ``tools/pygrub/src/GrubConf.py'' of Xen which could allow a user with ``root'' access in a guest operating system through specific crafted contents in grub conf to use domain 0 for running various commands at booting time of guest operating system. Fully functional interposition, inspection and complete isolation are not achieved in VMMM yet and need further investigation.  

\subsubsection{Backup Issues}

The sensitive data belonging to various business enterprises must be backed up by the SaaS providers to be used for fast recovery in disasters cases. Also, to protect against security threats like accidental leakage of data various encryption schemes be used to protect the back up data. These encryption schemes must be strong enough to resist modern attacks.

Amazon as cloud vendor does not encrypt the data by default at rest in S3. This control is given to the user to secure their back up data separately in order to protect against unauthorized access or tempering. Various tests can be performed to validate that a back up data is secure provided by SaaS model. These tests include; i) Storage insecurity and ii) Configuration insecurity. Any flaws identified by these tests may be potential threats which can lead unauthorized users to access information which is sensitive and stored in cloud backups belonging to different enterprises. 

\begin{table*}
	\centering
	\caption{Security solution for cloud storage architecture (Part-II)}
	\begin{tabular}{@{}p{3cm } |p{4cm} p{9cm}@{}}
	\toprule
	\textbf{Security Issue} &  \textbf{Solution} &  \textbf{Description} \\ 
	\midrule
	Data Dynamics & Public Auditing & Efficient and scalable public auditing system should be introduced to extend the existing schemes \\
    Data validity & 	Data Segregation & Security layers give you the flexibility to consolidate vast amounts of data while controlling who can see what,  through a sophisticated system of work groups, organizational rollups, and access levels, combined with field and function level security. \\
    Virtualization Vulnerabilities & Root Security     & Root protection enables users (e.g. smartphones, tablets)  with the android mobile operating system to get privileged control (known as root access) over different Android subsystems \\
    Backup Issues  & Encryption schemes  &  Different encryption schemes coded the data before storing. This secure the backup data from unauthorized users. \\
    \multirow{2}{*}{Availability}   & Multitier architecture  &  Multitier architecture (often referred to as multi-level architecture) or multi-layer architecture is a client-server structure in which the functions of presentation, application processing and data management are separated. \\
	 & Load Balancing 	      & Running on different servers, resilient to software and hardware failure, and be protected against DOS and DDOS attacks \\	
    Data Locality &  Regional backup servers  & Due to the different, region cyber rules, data should be kept in the same region servers to avoid data locality, cultural and cyber rules issues.   \\
	\bottomrule
	\end{tabular}
	\label{mgnSol}
\end{table*}


\subsubsection{Availability}

The SaaS applications guarantee around the clock services to a client. This involves architectural level changes in SaaS infrastructure and applications to attain availability and scalability. Multitier cloud architecture needs to be adopted, cloud architecture must also support load balancing of application instances, running on different servers. Cloud storage must be resilient to software and hardware failures further, it must be protected from both distributed denial of service attacks (DDOS) as well as denial of service DOS attacks\cite{p64,p65,p66,p67}.

For any unforeseen disaster, appropriate disaster recovery and operational sustainability action plan should be considered. This is important for certifying organizational data security and organizational nominal downtime. For example, at Amazon, the AWS API endpoints are hosted by the same world-class Internet infrastructure that Amazon supports and use connection throttling. To further reduce the potential impact of a DDOS attack, Amazon internally maintains the bandwidth that surfs on its vendor's internet bandwidth to validate the SaaS vendor's availability and evaluation tests.

Many applications automatically provide security locks for user accounts after successive incorrect credentials. Also, improper implementation and configuration of these functions can be vulnerable to malicious users as a result of DDOS attacks.   

\subsubsection{Data Locality}
In SaaS cloud model, a client uses the application provided by the SaaS and their own business data, but the client is unaware of storage location of the data in the cloud \cite{p33,p68}. This may lead to several issues and many cases. For example, due to data privacy laws in different counties, data locality is of utmost importance in enterprise business architecture. For instance in many Southern American States and several countries in European Union, certain types of data may not be allowed to leave the country premises because of the sensitivity of the information. Similarly local Government’s laws and jurisdiction issues may arise in case of any type of investigation \cite{p69}. A secure SaaS model may be capable to provide reliability to its clients at the consumer data locality.

\section{Cloud Storage Future and  Opportunities}
\label{Opportunities}

\begin{figure}
\centering
  \includegraphics[width=1\linewidth]{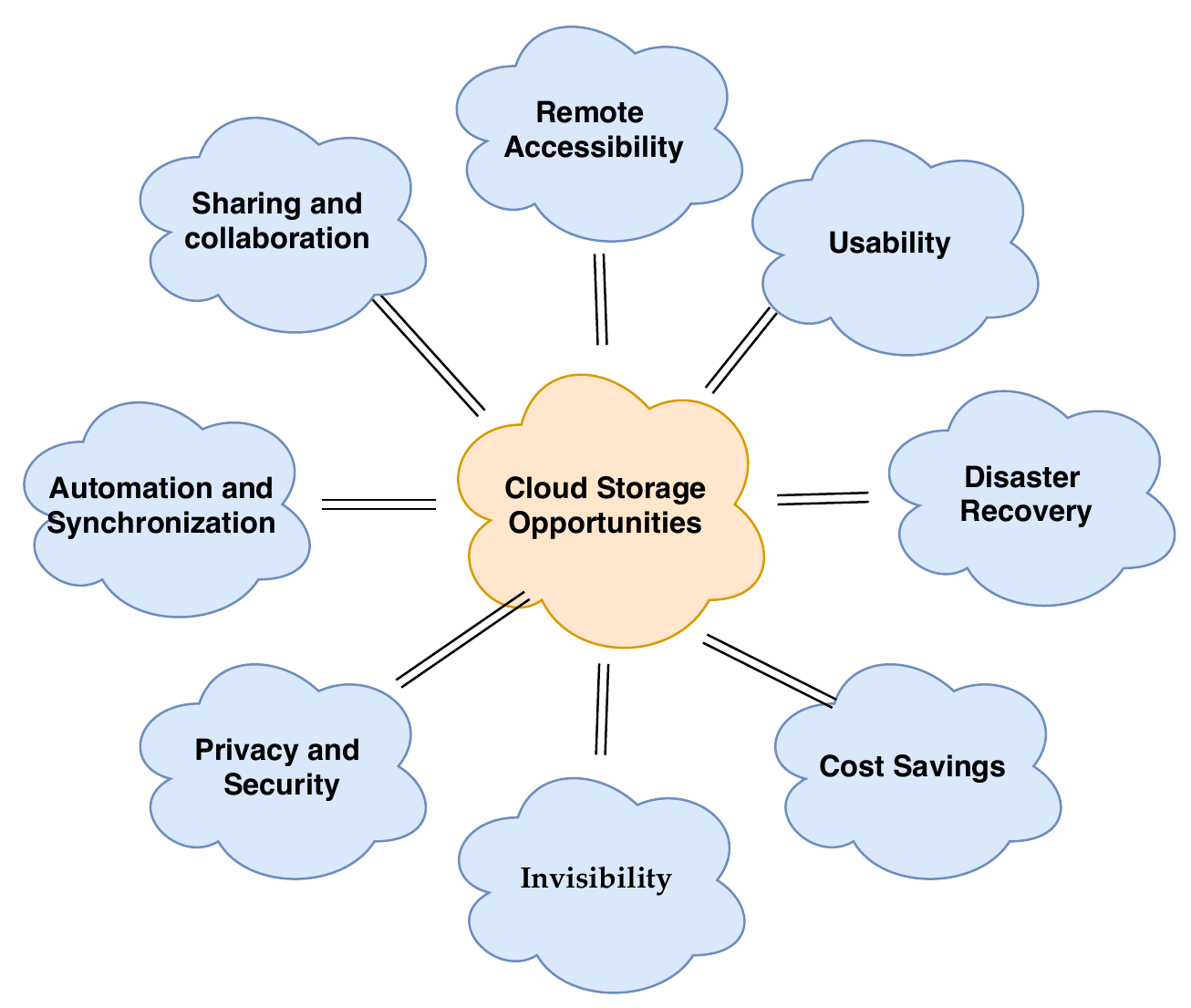}
  \caption{Opportunities of cloud storage}
  \label{fig:opportunities}
\end{figure}

The future of the cloud is not less than a dream. AI-enabled objects (e.g, self-driving vehicle), the web of IoT devices, and  5G  connectivity (i.e, mobile communication technology) is changing the way of living  \cite{Prop154}.  
The IT industry is rapidly changing everything. Its simple and easy user interface; no cost and capacity constraints; and other numbers of features are attracting the individual and market \cite{Prop155}.  The opportunities of cloud storage is listed in Fig \ref{fig:opportunities}.

The recent advances in smart technology generate a \emph{massive data} traffic. The 51 billion devices forecast is a big number; even seven times greater than the whole world population \cite{Prop149}. These devices will increase the annual size of the global data-sphere up to 175 ZB  \cite{Prop17,Prop154}.  Another report states, as shown in Figure \ref{fig:forecost},   that more than 331 billion dollars will be invested in cloud up to 2023 \cite{Prop17}. It needs special techniques and infrastructure to process the incoming data \cite{Prop04}. Furthermore, integration of  Artificial Intelligence (AI) in smart devices increases the data production value dramatically. With this rapid development in smart technology, cloud storage is getting more and more attention. Along with the AI, block-chain is adding safety and security to the cloud storage. This technology in storage is getting mature and will increase the customer trust on cloud. Furthermore, data compression is playing a good role in data archives by reducing the size of data storage on storage devices.  

This section presents a quick review of the cloud storage future and its   opportunities \cite{ad01,p33,ad03,ad04,ad05,ad06,ad07,ad08,ad09,ad10}.

\subsection{Remote Accessibility} 
Remote accessibility (i.e, access from everywhere and anytime) is the core of cloud storage. The fast network speed and AI is making it more smarter and faster.   Leading cloud provider (i.e, Apple iCloud \cite{Prop150}, Microsoft OneDrive \cite{Prop151}, and Google Drive  \cite{Prop152} etc) are providing fast and reliable remote services to their users.  
Remote access allows to store and retrieve items from a  cloud storage  without needing to create a physical connection. 
Accessibility of storage devices is getting interested after introducing high storage devices and high bandwidth network.  Remote access increases the usage of cloud storage and business. In the presence of internet services, cloud storage can provide seamless access to data files \cite{ad03}.  The coming 5G internet service will make the accessibility very easy and smart as real-time access \cite{Prop153}.

\subsection{5G Connectivity}  
With this high-speed technology,  humans will be able to virtually operate any machine at a distance of thousands of KMs \cite{Prop160}.  This will reduce the latency of up to 0 ms.  Such a big speed will minimize the need for the local hard drive. This technology will able to store and process data on the cloud without facing any jitters or delay.  It is making real-time use possible. 5G is a new era of cloud storage \cite{Prop156,Prop157}. 

\subsection{Internet of Things (IoT)}
With the introduction of the Internet of Things (IoT), the number of devices connected to the internet has increased enormously. By 2025, 75 billion devices are expected to be connected to the Internet processing 75 ZB data annually. This is a great number and will need a high technology to process and store this data. These figures clearly shows that the cloud storage has great worth in coming years. Furthermore, the use smart devices are also dramatically increasing. These  devices are small in size and have not enough space to store or process big data therefore, they depend on cloud \cite{Prop159}.

\subsection{Artificial Intelligence (AI)}
From facial expressions to self-driving vehicles, AI is progressing very rapidly. AI is making smart decisions in complex situations.  The today AI is called the weak AI which performs limited tasks such as recognizing facial expression and driving a car, however, the future will have general AI which will perform a task just like human beings. The AI is making the cloud storage further smarter and attractive \cite{Prop158}. Furthermore, the use of block-chain in storage is making it more secure \cite{Prop164}.

\subsection{Usability } 
The provider business directly depends on resources utilization. Today technologies massively increase the cloud usage because it provides a very easy and reliable user interface. Usually, cloud storage has a local desktop folder for PCs and mobile devices which allows users to move files back and forth between the cloud and the local system using drag and drop facilities \cite{p33,ad03,ad07}. 
The integration of smarts technologies (i.e, IoT, AI, fog and 5G), making the cloud storage usability very easy. The 5G will provide a high bandwidth like real-time access. Its cost is very low compared to buy the devices; which is very appealing \cite{Prop153}.

\subsection{Disaster Recovery}
In today modern world, data is the most valuable asset. Losing it, cause irreversible damage to the business (including loss of productivity, income, reputation and even customers). 
Business enterprises use cloud storage as a backup for their important files. In cloud storage, data is stored in three different locations and in case of any disaster, data may easily be recovered.  Furthermore,  cloud storage provides remote access to files therefore, these files can be used for recovery of their system in case any emergency or disaster \cite{p33}. 5G technology made the recovery process very easy and fast. Comparatively to the traditional disaster recovery, cloud storage recovery is very easy, cheaper and fast.  High investment, staff and maintenance are required for local disaster recovery site . 

\subsection{Cost Savings}
When we talk about cloud, it means that we are getting the resources of a supercomputer at our home without buying it. We actually, hire these resources on very cheaper rates which save the capital investment of the consumer. 
Cloud storage is used by various types of business enterprises to reduce their annual database operation expenses. Especially, the medium corporations, which are not able to invest too much on storage infrastructure, hire the cloud storage. This saves their major investment.  Storing one gigabyte of data using cloud storage services cost about three whereas a user can achieve further saving in terms of power consumption as remote cloud storage does not need internal power \cite{ad08,ad09}.  Cloud storage saves operational and maintenance cost and just as per their usage. 
 
\subsection{Invisibility} 
The word storage create the imagination of a big physical device to store big data.  Big data and storage mean a big physical device, which will need operation and maintenance. However, cloud storage does not need physical space and user access it remotely \cite{Prop162}. 
Cloud storage services, use virtualization techniques to provide resources to the customers. Customers do not know the complexities and working of the back end. Cloud storage is invisible and provides storage transparency, with no physical presence on the user side. It does not take up valuable space in the office or at home.  It does not need to spare a huge space for rocks and storage. Customers only hire the services and use them on the go. 
	
\subsection{Privacy and Security}
Security of cloud storage for sensitive and confidential information is usually higher than that for the locally stored data, especially for enterprises.  It uses advance security (i.e, advanced firewalls, event logging, internal firewalls, intrusion detection, data replication,  encryption,  and physical security) to protect the data from outside attacks.  Different type of security layers is used to protect the data houses. Concerning individual storage, enterprises invest more in security. Storage services in the cloud used encrypted data both in transmission as well as at rest ensuring no unauthorized access to data files.  AI, 5G, IoT and block chain are improving the privacy and security \cite{Prop161}. 
	
\subsection{Automation and Synchronization}
Cloud automation is a  term for the processes and tools that are used to reduce the manual effort involved in provisioning and managing cloud workloads. The cloud storage is self managed and does not need any human efforts  \cite{Prop163}. 
The main issue most businesses and customers have, the proper follow up of data backup. Cloud storage provides an automated data backup service to ease this tedious process. A user simply needs to tell the system what and when to back up, and the cloud service takes care of it by itself.      

Another attraction with the cloud storage is automatic synchronization. Synchronization process ensures that user data files are automatically updated across all of the user devices. In this way the latest versions of the user's data files are saved on his/her local device and available on all of other user devices like user Smartphone etc. 5G made the syncing more easy and now the devices works on real time \cite{Prop163}.       
	
\subsection{Sharing and collaboration} 
Cloud storage makes the sharing easy. Either it is a photo or a file or even a folder containing hundreds of information files, storage service in cloud make it convenient for a user to share it with a few clicks. Furthermore, it makes the files availability everywhere and every time \cite{Prop166}.       
Online cloud storage services are also ideal for collaboration purposes. It allows multiple users to collaborate and edit on a single document or data file. User do not have to concern about tracking the up-to-date version or who has made what changes \cite{ad03}.

\subsection{Massive devices and data}
As mentioned earlier, up to 2025, approximately 75 billion devices will connect to the internet and this will process more than 175 ZB of data per year. This is a very large figure and requires a lot of cloud storage. These predictions will drastically change the need for cloud storage. This shows that cloud storage has a very bright future ahead.

\section{Conclusion and future directions}
\label{conclu}

 The recent advances in IT industry  (e.g, Cloud computing, Internet of Things (IoT), Fog computing,  Artificial Intelligence (AI) and Block-chain) is rapidly revolutionizing the cloud storage. Especially, the 5G facilitation  (i.e, minimum access delay and ultra high speed) boost the use of cloud storage dramatically. This article presented different challenges, their counter measures, opportunities and future of cloud storage. it seems that cloud storage is designed to be highly scalable and conveniently manageable storage system rather than an efficient file system.   

Further, it is revealed that despite the ease of use and economic benefits, cloud storage technology still suffers from numerous problems.  The cloud storage architecture is mostly clouded by security  (e.g, confidentiality, integrity, access, authentication, authorization and data breaches) and data management issues (e.g, dynamics, data segregation, backup, and virtualization). To counter these threats, various measures are proposed in the literature. For example, for the security of data, digital certificates are used along with a trusted timestamp approach. Similarly, confidentiality is ensured through cryptography solutions while for data access issues, attribute-based encryption is mostly used. Access control is achieved through authentication and authorization. To maintain the integrity of the data, a global transaction manager is used to ensure fail-safe management of transaction across multiple databases.

Finally, it can be concluded that cloud computing (along with the integrated technologies) is a fast-growing technology which rapidly changing traditional computing. However, still, a lot of research efforts are needed to attract customers, especially business and enterprise customers to store their sensitive data, using cloud storage.

\bibliographystyle{IEEEtran}
\bibliography{references}


 
\EOD

\end{document}